\documentclass[12pt]{article}
\usepackage{epsfig}
\usepackage{amsfonts}
\usepackage{amscd}
\usepackage{latexsym}
\usepackage{amsmath,amssymb,amsthm}
\usepackage{verbatim}
\usepackage{setspace}
\usepackage{color}
\usepackage{cite}
\usepackage{graphicx}
\usepackage{mathtools}
\usepackage[all]{xy}

\usepackage[textheight=9in, textwidth=6.5in, letterpaper]{geometry}

\usepackage{color}   
\usepackage{hyperref}
\hypersetup{
    colorlinks=true,  
    linktoc=all,     
    linkcolor=black,  
    citecolor=black,
    filecolor=black,
    urlcolor=black,
}

\numberwithin{equation}{section}

\def\p{\partial}
\def\cl{{\cal L}}

\def\<{\langle}
\def\>{\rangle}

\def\be{\begin{equation}}
\def\ee{\end{equation}}
\def\beq{\be\begin{array}{c}}
\def\eeq{\end{array}\ee}
\def\bes{\be\begin{split}}
\def\ees{\end{split} \ee}
\def\bs{\begin{split}}
\def\es{\end{split} }

\def\a{{\alpha}}

\def\e{{\epsilon}}

   \makeatletter
  \let\over=\@@over \let\overwithdelims=\@@overwithdelims
  \let\atop=\@@atop \let\atopwithdelims=\@@atopwithdelims
  \let\above=\@@above \let\abovewithdelims=\@@abovewithdelims
\renewcommand\section{\@startsection {section}{1}{\z@}%
                                   {-3.5ex \@plus -1ex \@minus -.2ex}
                                   {2.3ex \@plus.2ex}%
                                   {\normalfont\large\bfseries}}

\renewcommand\subsection{\@startsection{subsection}{2}{\z@}%
                                     {-3.25ex\@plus -1ex \@minus -.2ex}%
                                     {1.5ex \@plus .2ex}%
                                     {\normalfont\bfseries}}

\begin{document}
\begin{titlepage}
\unitlength = 1mm

\vskip 1cm
\begin{center}

{ \LARGE {\textsc{Phase Space on a Surface with Boundary via  Symplectic  Reduction}}}

\vspace{0.8cm}

Vyacheslav Lysov

\vspace{1cm}

{\it 
Okinawa Institute for Science and Technology \\
1919-1 Tancha, Onna-son, Okinawa 904-0495, Japan
}

\begin{abstract}
We describe the  symplectic reduction construction for the physical phase space in gauge theory and apply it for the  BF theory. Symplectic reduction theorem allows us to rewrite the same phase space as a quotient by the gauge group action, what matches with the covariant phase space formalism.    We extend the symplectic reduction method to describe the phase space of the initial data on a slice  with boundary.  We show that the invariant phase space   has description in terms  of  generalized  de Rham cohomology, what makes the topological properties of  BF theory manifest.  The symplectic reduction 
can be done in multiple steps using different decompositions of the gauge group  with interesting finite-dimensional   intermediate symplectic spaces.
\end{abstract}

\vspace{1.0cm}

\end{center}

\end{titlepage}

\pagestyle{empty}
\pagestyle{plain}

\pagenumbering{arabic}

\tableofcontents

\section{Introduction}

Recently there is a lot of interest in phase space construction for the slices with boundary. The applications involve phase  space of matter  fields for the  ``island" approach to the  information puzzle\cite{Penington:2019npb,Almheiri:2019psf,Almheiri:2019hni}, asymptotic symmetries and soft theorems \cite{Strominger:2013jfa,He:2014laa,Kapec:2014opa} and entanglement entropy computations \cite{Donnelly:2014fua,Donnelly:2015hxa}.  The common feature of such phase spaces is existence of the  degrees of freedom associated to the boundary, which require a proper inclusion into physical phase space. There are multiple different ways of doing so, yet the resulting phase spaces are remarkably similar.  

The most common approach is the covariant phase space formalism \cite{Wald:1999wa}. The phase space is identified with the space of classical solutions for the equations of motion. Such 
solutions can be labeled by the boundary data on Cauchy slice, given a uniqueness of the Cauchy problem.  The boundary data is constrained by  certain components of the  
equations of motion. The symplectic form on space of solutions, written in terms of the boundary data  is degenerate along the gauge transformation directions, so an extra projection on gauge orbits is required. 

The covariant phase space approach can be generalized to the slices with boundary. In that case we need to properly analyze the Cauchy problem on space-time with corners, to figure out the relevant initial data and constraints,  what  turns to be a hard problem. Alternatively, we can use some  physical arguments such as edge modes or asymptotic symmetries to modify the phase space. 
Unfortunately, such modifications rely on fine details of the problem at hand such as existence of a good gauge, reasonable guess on boundary conditions etc.

In this paper we propose a different approach to construction of the physical phase space: symplectic reduction. Physical phase space is identified with the symplectic reduction of the   bare phase space, the space of the boundary values of fields, with respect to the  gauge symmetry action. The symplectic reduction theorem allows us to rewrite the symplectic reduction as a quotient space by the gauge group action. The quotient representation for the phase space matches with the covariant phase space construction. 

The advantage of using the symplectic reduction approach  is due to the existence of the reduction in stages procedure. We can split the full gauge group of the theory into several subgroups and perform a consecutive reductions with respect to them. In particular, we can use the subgroup of gauge transformations trivial on the boundary to construct the phase space with asymptotic symmetry action.  Unfortunately, 
most  of the symplectic reduction results require finite-dimensional phase spaces.  To counter this issue we used the  BF theory as the prime example in our analysis. The BF theory is a topological theory with finite-dimensional invariant phase space. We can arrange the reduction in stages in a such a way that the interesting features such as edge modes and asymptotic symmetry can be realized on a finite-dimensional phase space. 

In absence of boundary the invariant phase space for BF theory has nice algebraic topology description in terms of de Rham cohomology, what makes topological properties of the theory manifest. We show that this feature of the phase space can be extended to the case with nontrivial boundary. The ordinary de Rham cohomology become modified to the mapping cone, relative or compactly supported 
de Rham cohomology depending on a particular choice reduction.

\section{Phase space in QFT} \label{phase_cft}
The modern approach to the QFT uses the path integral as a guiding principle, so in this section we want to describe the field theory phase space using the path integral data.
\subsection{Path integral}
Let us consider QFT on a  manifold $M$. We can  define a  space  of fields $F(M)$ on $M$ and describe the partition function $Z(M)$  for the theory with action $S: F(M) \to \mathbb{R}$ as the path integral  over $F(M)$
\be\label{path_int}
Z(M)  = \int_{F(M)} \mathcal{D} \mathcal{A}\; e^{i S(\mathcal{A})},\;\;\; 
\ee
For  $M$ with   nontrivial boundary $\Sigma=\p M$ we need to additionally  specify boundary conditions for the fields on $\Sigma$ and the  partition function $Z(M)$ acquires dependence on the 
boundary values of fields.  Furthermore the dependence is very special: partition function $Z(M)$  becomes an element of the Hilbert space $Z(\Sigma)$ associated with the boundary $\Sigma$
\be\label{part_function_boundary}
Z(M) \in Z(\Sigma).
\ee
The (\ref{part_function_boundary}) is well known in topological quantum field theory (TQFT)  as one of Atiyah axioms \cite{PMIHES_1988__68__175_0}. There are many more examples where (\ref{part_function_boundary}) holds, so it is reasonable to conjecture it  being the universal property of the path integral. The natural  question is: 
\begin{center} 
{\it How do we construct   $Z(\Sigma)$ from the path integral data? }
\end{center}
 The short  answer is that the  $Z(\Sigma)$ is constructed by quantization of the boundary phase space $(\mathcal{M}_\Sigma, \omega_\Sigma)$. Such approach works well for  Chern-Simons theory \cite{hitchin1990flat}, where the  boundary Hilbert space $Z(\Sigma)$ is a (geometric) quantization of the moduli space of flat connections $\mathcal{M}_\Sigma$. In a path integral description the  $\mathcal{M}_\Sigma$ is the space of the boundary value of fields, i.e the  pull back  $i_\Sigma^\ast F(M)$ of the configuration space $F(M)$ using a  natural embedding $i_\Sigma : \Sigma \hookrightarrow M$.  The symplectic structure $\Omega_\Sigma$ on $\mathcal{M}_\Sigma$  is encoded in a  path integral (\ref{path_int}) as follows:     An action in (\ref{path_int})\be
S(\mathcal{A}) = \int_M L(\mathcal{A})
\ee
 is  an integral of a Lagrangian density $L(\mathcal{A})$\footnote{We use calligraphic letters $\mathcal{A}, \mathcal{B} $ for the fields on $M$, capital letters $A, B$ for fields on $\Sigma$, the boundary  of $M$.  } over $M$.
 The variation of $L(\mathcal{A})$
\be\label{pre_potential}
\delta L(\mathcal{A}) = L(\mathcal{A}+\delta \mathcal{A}) -L(\mathcal{A}) = \mathcal{E}(\mathcal{A}) \delta \mathcal{A} +d \Theta (\mathcal{A}, \delta \mathcal{A})
\ee
can be rearranged into (Euler-Lagrange) equations  $\mathcal{E}(\mathcal{A})$ and a boundary term $\Theta (\mathcal{A}, \delta \mathcal{A})$.
We can pull back the boundary term to $\Sigma$
\be
\theta (A, \delta A) = i_\Sigma^\ast \Theta (\mathcal{A}, \delta\mathcal{A}) ,\;\;\; A =i_{ \Sigma^\ast }\mathcal{A} = \mathcal{A}|_{\Sigma}
\ee
and identify it with the symplectic potential while the symplectic form $\omega_\Sigma$ on $\mathcal{M}_\Sigma$ 
\be\label{sym_form_potential}
\omega_\Sigma = \int_{\Sigma} \delta \theta (A, \delta A).
\ee

\subsection{Gauge theories}

The path integral for the gauge theory is defined as an integration over the gauge invariant configurations.  
Two configurations related by a gauge group are  considered identical
\be
\Phi_g:  \mathcal{A} \mapsto   \mathcal{A}^g =g^{-1}\mathcal{A}g +g^{-1}dg,
\ee
with $g$ being an element of $G^M$, the  gauge group\footnote{In our discussion we use the math literature notation $G^\Sigma$ for the gauge group acting on $\Sigma$, what is very convenient  to distinguish the gauge groups associated to  $\Sigma$ and boundary $\p \Sigma$.}, a   space of maps from $M$ to a Lie group $G$ i.e.
\be
g\in G^M = \hbox{Maps} (M, G).
\ee

For  $M$ without boundary the  $Z(M)$ is a number, so   the gauge invariance of the path integral (\ref{path_int})  implies that the  action
is invariant under the gauge transformation. The invariance of the action implies that the   Lagrangian  $L(\mathcal{A})$ may change by a  total derivative
\be
L^g(\mathcal{A}) =  L ( \mathcal{A}^g)  = L(\mathcal{A}) +  d J(\mathcal{A}, g).
\ee
The corresponding symplectic potential density on $\Sigma$ changes by  the $\delta$-exact term
\be
 \theta^g(A, \delta A)  =\theta(A, \delta A)+ \delta J(A,g)
\ee
The symplectic form  $\omega_\Sigma$ defined in (\ref{sym_form_potential}) stays invariant, so   the boundary  gauge group  $ G^\Sigma$   action  is a symplectic symmetry of the boundary phase space $(\mathcal{M}_\Sigma, \omega_\Sigma)$.  

\subsection{Invariant phase space}

 The algebra of classical observables is the algebra of  functions on phase space  $C^\infty (\mathcal{M}_\Sigma)$.  The gauge invariant classical observables are 
$G^\Sigma$-invariant functions. We can describe the gauge invariant observables on $\mathcal{M}_\Sigma$ in terms of arbitary observables on a small symplectic space 
\be
 \mathcal{M}_\Sigma^{inv} = \mathcal{M}_\Sigma//G^\Sigma,
\ee
known as the {\it symplectic reduction of  $\mathcal{M}_\Sigma$ with respect to the action of $G^\Sigma$}. The symplectic reduction theorem allows us to describe the $ \mathcal{M}_\Sigma^{inv} $ as a quotient space
\be
\mathcal{M}_\Sigma^{inv} = \mu_G^{-1}(0)/G^\Sigma.
\ee
The  moment map  $\mu_G: \mathcal{M}_\Sigma \to \mathfrak{g}^{\Sigma \ast} $ takes  values in dual Lie algebra   $\mathfrak{g}^{\Sigma \ast}$  of $G^\Sigma$. The symplectic reduction realization of the  phase space in field theory is well known in case of  Chern-Simons theory \cite{Alekseev:1993rj}.

\section{Phase space construction in presence of boundary}\label{bound_pahse_sect}
Our phase space  construction from section \ref{phase_cft} provides a phase space for the manifold $\Sigma$, which is the boundary of the space-time manifold $M$. Being boundary implies that $\Sigma$ does not have the boundary itself i.e. $\p \Sigma = 0$.   Our goal is to generalize the invariant phase space construction to include  $\Sigma$ with boundaries.  Indeed, for the construction we only need $\Sigma$ to be part of the boundary $\Sigma \subset \p M$  of codimension $1$. Since $\Sigma$ has codimension 1, it is very natural to call it a hyperfurface, while we further will refer to it as a surface for simplicity.     The  generalization of the  construction from section \ref{phase_cft}  requires to deal with certain problems that we outline below.

\subsection{Infinite-dimensional phase spaces}

Most of the classical results in symplectic geometry, we review  in section \ref{sym_sec}, especially the  symplectic reduction theorem,  require finite-dimensional symplectic manifolds.  Unfortunately, the phase spaces we typically use in QFT are infinite-dimensional, so there could be  potential issues with  the symplectic reduction. In present paper we do not try to construct a theory of infinite-dimensional symplectic reduction,  but rather we hope that infinite-dimensional reduction works for one of the simples QFT: the BF theory.  Furthermore, the BF theory is known to be {\bf topological theory with finite-dimensional invariant phase space} for the compact $\Sigma$. We use the topological theory  features as a consistency check for infinite-dimensional symplectic reduction results with details described in sections 
\ref{BF_theory} and \ref{inv_phase_features}.

\subsection{Boundary terms in symplectic form}

Our definition (\ref{pre_potential}) of symplectic potential allows for an  arbitrary shift of the form 
\be
\Theta \to \Theta + dK,
\ee
so  the symplectic form (\ref{sym_form_potential}) is defined up to a possible boundary term 
\be
\omega_\Sigma = \int_{\Sigma} \delta \theta (A, \delta A)+ \int_{\p \Sigma} \delta K (A, \delta A).
\ee
The boundary terms in symplectic form play the key role in our discussion.  There are various ways to fix the boundary terms. One popular approach  \cite{Donnelly:2016auv}  is to require that the 
pre-symplectic potential is invariant under field-dependent gauge transformations.  We propose to {\bf use the non-degeneracy of $\omega_\Sigma$ to fix the boundary terms}. The details of our approach are presented in section \ref{bound_pairing}.
\subsection{Non-degenerate pairing}
By definition the symplectic form is a non-degenerate two form. In our analysis we want make the non-degeneracy manifest. In case of abelian BF theory the phase space is a linear symplectic space of the form $V\oplus W$, with vector spaces $V,W$ being differential forms of a certain degree.   In section \ref{sym_manif} we show how to  define the  symplectic form on $V\oplus W$   from a non-degenerate pairing $V\times W \to \mathbb{R}$. The pairing for BF theory  is the integration of the differential forms over  $\Sigma$  and requires a proper modification in presence of boundary, what we discuss in section \ref{bound_pairing}.  The gauge symmetries of BF theory    can also  be identified with the differential forms,  while the moment map  requires the  dual to the  gauge symmetry algebra. The dual algebra can  be constructed using the same differential form pairing.

\subsection{Central extension of the symmetry algebra}
The Poisson bracket realization of phase space symmetries allows for the central extension, what we briefly review in section \ref{sym_sect}. The phase space  for the surface with boundary requires 
edge mode inclusion, what leads to additional symmetries and centrally extended symmetry algebra. This feature was observed in various situations like Chern-Simons theory   \cite{Alekseev:1993rj},
general relativity \cite{Compere:2018aar} and other theories.     In section \ref{BF_symm_central} we show that the BF theory also has centrally-extended algebra of symmetries. In case of centrally extended symmetry the symplectic reduction theorem requires a generalization, which we briefly review in sections  \ref{non_equiv_reduct}, \ref{red_stages} and implement for BF theory in section \ref{bf_red_stages}. 
  
\subsection{Duality in BF theory}
The abelian BF theory of $p$-form  in $d+1$-dimensional space, at least on the classical level, is the same as BF theory of $(d-p)$-form  on the same space. This property follows from the integration by parts 
\be
S_{BF}^{(p)} = \int_{M} d\mathcal{A} \wedge \mathcal{B}    =(-1)^{1+p(d-p)}  S^{(d-p)}_{BF}+ \int_{\p M} \mathcal{A} \wedge \mathcal{B}. 
\ee
The symplectic form $\omega_\Sigma$, defined from the action above, is independent on the boundary terms, so both $p$-form and $(d-p)$-form theories have the same phase space. Indeed, an explicit constriction in section \ref{BF_theory} shows that phase space is manifestly symmetric under the $p\to d-p$ exchange. In a presence of boundary, when we add edge modes and modify symmetries, it is far from obvious that the invariant phase space   remains invariant under  the duality.   In a literature \cite{wu1991topological} the proposed phase  space for BF theory does not have this symmetry, while we show in section \ref{inv_phase_features} that there is a phase space   with such  symmetry.

\section{Symplectic geometry review} \label{sym_sec}
The symplectic reduction requires certain terminology from symplectic geometry, which we briefly review in this section. 
\subsection{Symplectic manifolds}\label{sym_manif}
A pair $(\mathcal{M}, \omega)$ defines a {\it symplectic} manifold, if $\mathcal{M}$ is a smooth manifold, endowed with   a non-degenerate closed two-form $\omega$. 

There is a particular type of the symplectic manifolds,  useful for our analysis, constructed from a vector space and its dual.  Given a   finite-dimensional vector space $V$, the vector space $\mathcal{M}= V \oplus V^\ast$  admits a canonical symplectic structure 
\be\label{lin_sym_space_form}
\omega_0 ((v_1,\alpha_1), (v_2, \alpha_2)) = \alpha_2 (v_1) - \alpha_1 (v_2) = \<\alpha_1, v_2\> -\<\a_1, v_2\>,
\ee
where we used standard notation for the canonical pairing 
\be
\<\cdot, \cdot\>: V \oplus V^\ast  \to \mathbb{R}.
\ee
The  pair $(\mathcal{M}, \omega_0)$ defines a symplectic manifold with linear structure often referred as {\it linear symplectic space}. 

In our analysis we need an infinite-dimensional generalization of the construction above. Given a pair of infinite-dimensional vector spaces $V$  and $W$ and a non-degenerate pairing between them 
\be
\<\cdot, \cdot\>:\; V \oplus W \to \mathbb{R}. 
\ee
 we can  define the symplectic form on $\mathcal{M} = V\oplus W$  to be 
\be
\omega ((v_1,w_1), (v_2, w_2)) = \<w_1, v_2\> -\<w_1, v_2\>.
\ee
We can write $\omega$ as a 2-form on $\mathcal{M}$\footnote{Here we are  using $\delta$ for differential on the field space to reserve the usual    $d$  for the  differential on a space-time. }
\be\label{linear_can_symp_form}
\omega  =\< \delta v , \delta w \>. 
\ee

\subsection{Symmetries}\label{sym_sect}
  A smooth map $\Phi: \mathcal{M}\to \mathcal{M}$ is called a {\it symplectomorphism} (or a {\it canonical transformation}) if it is a diffeomorphism and it preserves the symplectic form i.e.
\be
\Phi^\ast \omega= \omega.
\ee 
An infinitesimal version of symplectomorphism, is a vector field $\xi$ such that  
\be
0=\cl_\xi \omega =( \iota_\xi \delta +\delta \iota_\xi) \omega  = \delta (\iota_\xi \omega).
\ee
Such vector fields form a Lie algebra $sym(\mathcal{M})$ with bracket being the vector field bracket. Locally, we can write any closed form as an exact form 
\be
\iota_\xi \omega  = \delta H_\xi.
\ee
 We can define the {\it Hamiltonian vector field} as the vector field $\xi$ such that  there is a global function $H_\xi$.  Hamiltonian vector fields form a Lie algebra $ham(\mathcal{M})$, which is subalgebra of all vector fields on $\mathcal{M}$. Similarly, we can turn  smooth functions $C^\infty(\mathcal{M})$ on $\mathcal{M}$ into Lie algebra with   {\it Poisson bracket} defined as 
\be
\{H_\xi,H_\eta\} = \iota_{\xi}\iota_{\eta} \omega.
\ee
The  Lie algebra $C^\infty(\mathcal{M})$ in general is a central extension of $ham(\mathcal{M})$ i.e. 
\be\label{central_ext_ham}
\{H_\eta, H_\xi\} = H_{[\eta, \xi]} +c(\eta, \xi).
\ee
 The simple example of central extension uses $\mathcal{M} = \mathbb{R}^2 $ with coordinates $p,q$ and canonical symplectic form 
$\omega  =\delta p\wedge \delta q$. The  pair of  commuting vector fields $\p_p$ and  $\p_q$  are Hamiltonian vector fields
\be
H_{\p_p} = q,\;\; H_{\p_q}  =- p,
\ee
while the  Poisson bracket between them is nonzero
\be
\;\; \{H_{\p_p}, H_{\p_q}\} = 1\neq 0  = H_{[\p_p,\p_q]}.
\ee

\subsection{Lie group action}
The  action of a  Lie group $G$  on a  symplectic manifold  $(\mathcal{M}, \omega)$ is a collection of symplectomorphisms 
\be
\Phi: G \to \hbox{Diff} (\mathcal{M}): g\mapsto \Phi_g,\;\;  \Phi_g: \mathcal{M}\to \mathcal{M},\;\;  \Phi_h \circ \Phi_g  = \Phi_{gh},\;\;  \Phi_g^\ast \omega  =\omega.
\ee
An infinitesimal version of the Lie group action  is a Lie-algebra morphism 
\be
V: \mathfrak{g} \to sym(\mathcal{M}): \;\; \e \mapsto V(\e),\;\; V([\e,\eta]) = [V(\e), V({\eta})],
\ee
that we will call  {\it weakly  hamiltonian action} if the image of $V$ is in $ ham(M,\omega) \subset sym(\mathcal{M},\omega)$.
The  {\it moment map}  $\mu$  for weakly  hamiltonian action $G$ on $(\mathcal{M}, \omega)$  is a smooth map $\mu: \mathcal{M} \to \mathfrak{g}^\ast$ such that 
\be\label{moment_map}
\<\xi,\mu \>  = H_{V(\xi)},\;\; \forall \xi \in \mathfrak{g},
\ee
where $\<\cdot , \cdot\>$ is the canonical paring between $\mathfrak{g}$ and $\mathfrak{g}^\ast$. The dual Lie algebra $\mathfrak{g}^\ast$, is equipped with the  canonical coadjoint action $Ad_{g^{-1}}^\ast$ of $G$, so we can define the  {\it equivariant moment map} as a moment map which obeys
\be
\mu (\Phi_g(x)) = \hbox{Ad}_{g^{-1}}^\ast \mu(x),\;\; \forall g\in G.
\ee 
The action of a Lie group $G$ on $(\mathcal{M},\omega)$ is called {\it Hamiltonian action} if there exists an equivariant moment map for this action. An  existence of  equivariant  moment map is equivalent to the absence of central extension in Poisson algebra (\ref{central_ext_ham}).

\subsection{Symplectic reduction}\label{symp_red_section}
The Hamiltonian action of a Lie group $G$ on a symplectic manifold $(\mathcal{M}, \omega)$ allows us to define new symplectic manifold 
 $(\mathcal{M}^{red}, \omega^{red})$, the   {\it symplectic reduction} of $\mathcal{M}$ by $G$ and denoted using $//$ notation  i.e
 \be
 \mathcal{M}^{red} = \mathcal{M}//G.
 \ee
The Marsden-Weinstein theorem \cite{marsden2007hamiltonian} provides us with explicit construction of $( \mathcal{M}^{red}, \omega^{red})$ in the form of quotient space 
\be 
\mathcal{M}^{red}= \mu^{-1}(0)/G.
\ee
The  symplectic form on $\mathcal{M}^{red}$ 
\be
 \omega^{red} =s^\ast i_\mu^\ast \omega 
\ee
is expressed in terms of  canonical embedding $i_\mu: \mu^{-1}(0)\hookrightarrow \mathcal{M}$  and  a  section $s: \mathcal{M}_{red} \to \mu^{-1}(0)$ of a principal G-bundle $\pi: \mu^{-1}(0)\to \mathcal{M}^{red}$. Furthermore the  $\omega^{red}$ is  independent of choice of a section.

\subsection{Non-equivariant symplectic reduction} \label{non_equiv_reduct}
 We can relax an assumption of the moment map being equivariant and define {\it non-equivariant symplectic reduction}.  We can  define 
the measure of non-equivariance by 
\be
c_g= \mu (\Phi_g(x)) - Ad_{g^{-1}}^\ast \mu(x)
\ee
 and  use it to  define the  {\it affine coadjoint action}
\be\label{affine_action}
G^a: \xi \mapsto g \cdot^a \xi   = Ad_{g^{-1}}^\ast \xi+c_g.
\ee 
The moment map $\mu$ becomes equivariant with respect to the affine action, so we can modify the quotient space description
 \be
 \mathcal{M}^{red} = \mu^{-1}(0)/ G_0^a,
 \ee
where 
\be
G_0^a = Stab_{G^a}(0)
\ee
is the stabilizer subgroup of $0\in \mathfrak{g}^\ast$ under the affine action (\ref{affine_action}).
\subsection{Reduction by stages}\label{red_stages}
Let $G$ and $K$ be two Lie groups with hamiltonian action on a symplectic manifold $(\mathcal{M}, \omega)$ with commuting actions. Then the following relation holds 
\be
\mathcal{M}// (G\times K) =( \mathcal{M}//G)//K = (\mathcal{M}//K)//G.
\ee 
There is generalization of the reduction by stages to the case of non-commuting actions which we can conjecture to be 
\be
\mathcal{M}// (G\times K) =( \mathcal{M}//G)//K^a_0 = (\mathcal{M}//K)//G^a_0
\ee 
were  affine actions 
\be
G^a:\eta \mapsto g \cdot^a \eta   = Ad_{g^{-1}}^\ast \eta+c^K_g,\;\;\; K^a:\e \mapsto k \cdot^a \e  = Ad_{k^{-1}}^\ast \e+c^G_k,
\ee
are defined from the non-equivariance 
\be
c^K_g(x) = \mu_K (\Phi_g(x)) - Ad_{g^{-1}}^\ast \mu_K(x)\neq 0,\;\; c^G_k(x) = \mu_G (\Phi_k(x)) - Ad_{k^{-1}}^\ast \mu_G(x)\neq 0.
\ee

Let us illustrate the conjecture in case of the symplectic reduction of the finite-dimensional linear symplectic space. Let $\mathcal{M}$ be an $2N$-dimensional linear symplectic space such that 
\be
\mathcal{M} = \mathbb{C}[e_1,f_1,e_2,f_2,...,e_N, f_N],\;\; \omega  = \sum_{j=1}^N \delta e_j\wedge \delta f_j.
\ee
Let us further chose the basis such that symplectic actions of  $n$-dimensional group $G$ and  $k$-dimensional group $K$ are of the form 
\be
G: \delta e_i = \e_i,\;\;\; i=1,..,n,
\ee
and 
\be
K: \delta e_{i+n} = \lambda_{i+n},\;\;\; i = 1,...,k-m;\;\;\; \delta f_i =- \lambda_{k-m+i},\;\;\; i=1,...,m.
\ee
The individual actions of $G$ and $K$ are hamiltonian  with equivariant moment maps 
\be
\mu_G  = ( f_1,...,f_n),\;\;\; \mu_K = (e_1,...,e_{m},f_{n+1},..., f_{n+k-m}), 
\ee
but the actions do not Poisson commute i.e.
\be
c^K_g(x) = \mu_K (\Phi_g(x)) - Ad_{g^{-1}}^\ast \mu_K(x) =(\e_1,...,\e_m,0,..,0) \neq 0,
\ee
\be
 c^G_k(x) = \mu_G (\Phi_k(x)) - Ad_{k^{-1}}^\ast \mu_G(x)  = (-\lambda_{k-m+1},...,-\lambda_k,0,...,0)\neq 0.
\ee
The stabilizer group of the affine action is 
\be
(G\times K)^a_0:\;\; \delta e_i = \e_i,\;\; i=m+1,..,n ,\;\;\;  \delta e_{i+n} = \lambda_i,\;\; i=1,... ,k-m,
\ee
while the reduced phase space 
\be\label{lin_red_ex}
\mathcal{M}// (G\times K) = (\mu_k^{-1}(0) \cap \mu^{-1}_G(0)) /(G\times K)^a_0 = \mathbb{C}[e_{n+k-m+1},f_{n+k-m+1},...,e_N, f_N ].
\ee
The reduction of $\mathcal{M}$ under the $G-$action is 
\be
\mathcal{M}//G = \mu^{-1}_G(0)/ G = \mathbb{C}[e_{n+1},f_{n+1},...,e_N, f_N],
\ee
while the $K$-action on reduced space is 
\be
K: \delta e_{i+n} = \lambda_{i+n},\;\;\; i = 1,...,k-m,
\ee
which is identical to the $K^a_0$ action on $\mathcal{M}$.  The $K$-action on $\mathcal{M}//G$ is hamiltonian with equivariant moment map 
\be
\tilde{\mu}_K = (f_{n+1},...,f_{n+k-m}).
\ee
The reduction in stages becomes 
\be
(\mathcal{M}//G)//K^a_0 = \tilde{\mu}^{-1}_K(0)/ K = \mathbb{C}[e_{n+k-m+1},f_{n+k-m+1},...,e_N, f_N],
\ee
which identical to the reduced phase space (\ref{lin_red_ex}). The reduction in stages 
\be
(\mathcal{M}//K)//G^a_0
\ee
can be performed in similar way with the end-result being identical to the (\ref{lin_red_ex}) as well.

\section{BF-theory}\label{BF_theory}

The abelian $p$-form BF theory on $(d+1)$-dimensional manifold $M$ is a field theory with field space 
\be
F(M)=  \Omega^{p}(M, \mathbb{R})\oplus  \Omega^{d-p} (M, \mathbb{R})
\ee
being the  space of differential forms. Instead  of usual  $\Omega^{p}(M, \mathbb{R})$ notation for real-valued $p$-forms on $M$ we will use the simplified notation $\Omega^p (M)$, since all differential forms in our discussion are real-valued.

The action for the BF-theory  
\be\label{BF}
S[\mathcal{A}, \mathcal{B}] = \int_M d\mathcal{A}  \wedge\mathcal{B}
\ee 
is invariant under the two types of gauge transformations 
\be\label{BF_gauge_tr}
G^M: \mathcal{A} \to \mathcal{A} +  d\e,\;\; \e \in \Omega^{p-1}(M),\;\;\; K^M: \mathcal{B} \to \mathcal{B}+d\lambda,\;\; \lambda\in \Omega^{d-p-1}(M).
\ee
There are many advantages in using the BF theory as a prime example. It is  defined in arbitrary dimension and  for arbitrary manifold $M$. The BF action (\ref{BF}) is metric-independent, what makes it  into a topological theory. The  gauge symmetries make the invariant phase space finite-dimensional.

\subsection{Phase space}
The symplectic space for  the theory (\ref{BF}) is  the space of boundary values of fields 
\be\label{M_BF}
\mathcal{M}_\Sigma  =  i_\Sigma^\ast F(M) =\Omega^p(\Sigma) \oplus \Omega^{d-p}(\Sigma)
\ee
with symplectic form 
\be\label{bare_sympl_BF}
\omega_\Sigma=- \int_\Sigma \delta A\wedge \delta B.
\ee
For compact $\Sigma$, i.e. $\p \Sigma=0$, the symplectic form (\ref{bare_sympl_BF}) is the   the canonical symplectic form (\ref{linear_can_symp_form}) on a  linear symplectic space
with a pairing being the Poincare pairing 
\be
\<\cdot,\cdot\>: \Omega^p(\Sigma) \times \Omega^{d-p}(\Sigma) \to \mathbb{R}:\;\; (A, B) \mapsto  \int_\Sigma A\wedge B.
\ee

\subsection{Invariant phase space}\label{sect_invariant_BF}

The infinitesimal  version  of gauge transformations  (\ref{BF_gauge_tr}) for the boundary fields 
\be\label{BF_gauge_no_bound}
\delta_g A = d\e,\;\; \delta_g B =d\lambda,\;\; \e \in \mathfrak{g}^\Sigma =\Omega^{p-1}(\Sigma),\;\; \lambda\in  \mathfrak{k}^\Sigma = \Omega^{d-p-1}(\Sigma)
\ee
are generated by the symplectic vector fields, i.e.
\be
i_{V(\e)}\omega_\Sigma =  -\int_\Sigma d\e \wedge \delta B =  (-1)^{p+1} \delta \int_\Sigma \e \wedge d B,
\ee
\be
i_{V(\lambda)}\omega_\Sigma = (-1)^{p(d-p)} \int_\Sigma  d\lambda \wedge \delta A =(-1)^{(p+1)(d-p) }   \delta \int_\Sigma  \lambda  \wedge dA.
\ee
The generating    Hamiltonians  on $\mathcal{M}_\Sigma$ are 
\be
H_{\e} = (-1)^{p+1} \int_{\Sigma}  \e  \wedge dB,\;\;  H_{ \lambda} = (-1)^{(p+1)(d-p) }  \int_{\Sigma}  \lambda \wedge dA.
\ee
Using Poincare duality we can express the dual Lie algebras 
\be
\mathfrak{g}^{\Sigma\ast } = \Omega^{p-1}(\Sigma)^\ast  = \Omega^{d-p+1} (\Sigma),\;\; \mathfrak{k}^{\Sigma\ast}  = \Omega^{d-p-1}(\Sigma)^\ast  = \Omega^{p+1} (\Sigma),
\ee
as differential forms, so the canonical pairing becomes the pairing on differential forms 
\be
\<\cdot, \cdot\> :  \mathfrak{g}^{\Sigma} \times \mathfrak{g}^{\Sigma\ast}   \to \mathbb{R}:\; ( \e,F) \mapsto \< \e,F\>=\int_{\Sigma} \e \wedge F.
\ee
By definition of the moment map
\be
H_{\e}  = \<\e, \mu_{G}\>,\;\;\; H_{\eta}  = \<\eta, \mu_{K}\>,  
\ee
we can construct
\beq\label{bf_moment_no_bound}
\mu_{G}: \mathcal{M}_\Sigma \to\mathfrak{g}^{\Sigma \ast} :   \Omega^{p}(\Sigma)\oplus \Omega^{d-p}(\Sigma) \to \Omega^{d-p+1}(\Sigma):\; (A,B) \mapsto  (-1)^{p+1} dB \\
\mu_{K}: \mathcal{M}_\Sigma \to\mathfrak{k}^{\Sigma \ast} :   \Omega^{p}(\Sigma)\oplus \Omega^{d-p}(\Sigma) \to \Omega^{p+1}(\Sigma):\; (A,B) \mapsto   (-1)^{(p+1)(d-p) }   dA.
\eeq
Moment maps in (\ref{bf_moment_no_bound}) are equivariant. Equivalently the Poisson algebra  of symmetries is trivial
\be
\{H_{\e}, H_{\eta}\}=0,
\ee
so we can use the symplectic reduction theorem to describe the  invariant phase space as
\be
\mathcal{M}^{inv}_{\Sigma} = \mathcal{M}_\Sigma //(G^\Sigma \times K^\Sigma)= \frac{\mu_K^{-1}(0) \cap \mu_G^{-1}(0)}{ G^\Sigma \times K^\Sigma}.
\ee
The symplectic reduction can be carried explicitly in the form of de Rham cohomology 
\be\label{BF_inv_no_bound}
\mathcal{M}^{inv}_\Sigma =  \frac{Z^{p}(\Sigma)\oplus Z^{d-p}(\Sigma)}{ d\Omega^{p-1}(\Sigma)\oplus d \Omega^{d-p-1}(\Sigma)} = H^{p}(\Sigma) \oplus H^{d-p}(\Sigma).
\ee
 The pullback of $\omega_\Sigma$ onto $\mu_{G} = \mu_K =0$,
\be\label{BF_pull_no_bound}
i_\mu^\ast \omega_\Sigma  =- \int_\Sigma \delta A\wedge \delta B,
\ee
is well-defined on gauge orbits, i.e. it is the same for $A$ and $A+d\e$. The choice of section $s:  \mathcal{M}^{inv}_\Sigma \to  \mu_K^{-1}(0) \cap \mu_G^{-1}(0)$ is the same as choice of gauge fixing.   
 The reduced symplectic form  is independent on choice of section 
\be\label{BF_form_no_bound}
\omega^{inv}_\Sigma =s^\ast i_\mu^\ast \omega_\Sigma =- \int_\Sigma \delta A\wedge \delta B,\;\; A \in H^{p}(\Sigma),\;\; B \in H^{d-p}(\Sigma).
\ee
The invariant symplectic form  is a canonical symplectic form (\ref{linear_can_symp_form}) on a linear symplectic manifold 
with pairing being the    Poincare paring for de Rham cohomology
\be
\<\cdot, \cdot\>: H^p(\Sigma) \times H^{d-p}(\Sigma) \to \mathbb{R}.
\ee

\subsection{BF theory as a topological theory}

The invariant phase space (\ref{BF_inv_no_bound}) is described in terms of well known mathematical objects: de Rham cohomology groups $H^p (\Sigma)$. The de Rham cohomology are known to be finite-dimensional, what makes $\mathcal{M}^{inv}_\Sigma$  into finite-dimensional phase space.  The diffeomorphism $f: \Sigma\to \Sigma$, homotopic to the identity, induces the isomorphism 
$f^\ast : H^p (\Sigma) \to H^p (\Sigma)$  of cohomology groups.  The  phase space   (\ref{BF_inv_no_bound}) depends only on topology of $\Sigma$, as expected in topological theory.

The infinitesimal version of a diffeomorphism $f: \Sigma\to \Sigma$ is vector field $v$ on $\Sigma$. The infinitesimal  transformation of $A$ is 
\be \label{inf_diffeo}
\delta_v A  = \cl_v A = d \iota_v A  + \iota_v dA
\ee
For $A\in H^p(\Sigma)$ the second term in (\ref{inf_diffeo}) vanishes, since $A$ is a closed form, while the first term is the shift by a total derivative, which is an equivalence relation 
in quotient space construction of the   $H^p(\Sigma)$.

\subsection{Covariant phase space}
Let us describe the  BF-theory phase space  by the means of covariant phase space construction.\footnote{Good introduction to this method you can find in \cite{Compere:2018aar}} The solutions to equations of motion 
\be
d \mathcal{A} =d\mathcal{B}=0
\ee
are  parametrized by boundary values 
\be
A = i_\Sigma^\ast \mathcal{A} =  \mathcal{A}|_{\Sigma},\;\;\; B =i_\Sigma^\ast \mathcal{B}=  \mathcal{B}|_{\Sigma},
\ee
subject to the constraint equations 
\be
i_\Sigma^\ast (d \mathcal{A} ) = d (i_\Sigma^\ast \mathcal{A}) =  dA =  dB=0.
\ee
Note that the constraint equations are identical to the $\mu_{G} = \mu_{K}=0$.
Thus the solution space, under the  uniqueness of Cauchy problem, is 
\be
{\mathcal Sol} (M) = Z^p(\Sigma)\oplus Z^{d-p}(\Sigma).
\ee 
The  pre-symplectic form on solution space 
\be
\omega_\Sigma^{cov}  = -\int_{\Sigma} \delta A \wedge \delta B,
\ee
 is degenerate due to   (boundary) gauge transformations 
\be
\delta A = d \e,\;\;\; \delta B = d\lambda.
\ee
The pre-symplectic form $\omega_\Sigma^{cov} $ is identical to the pull back of $i_\mu^\ast \omega_\Sigma$ in (\ref{BF_pull_no_bound}).
Gauge orbits are naturally parametrized by cohomology groups, so the gauge-invariant solution space 
\be
{\mathcal Sol}^{inv} (M) = H^p(\Sigma)\oplus H^{d-p}(\Sigma)
\ee 
is identical to the invariant phase space (\ref{BF_inv_no_bound}) we constructed using the symplectic reduction.

\subsection{Surface with boundary}
De Rham cohomology groups  $H^p(\Sigma)$ are well defined for $\Sigma$ with nontrivial boundary, so we can try to use the  (\ref{BF_inv_no_bound}) in a presence of boundary.  Unfortunately,  the symplectic form  (\ref{BF_form_no_bound}) becomes degenerate. We can  immediately see the degeneracy if we  consider $p=0$ BF theory where symplectic paring is a map 
\be\label{paring_candidate}
H^0(\Sigma) \times H^d (\Sigma) \to \mathbb{R}.
\ee 
The homology $H_p(\Sigma)$ are canonically dual\footnote{We are working over $\mathbb{R}$ so $H^p(\Sigma) = Hom (H_p (\Sigma), \mathbb{R}).$} to the cohomology groups, while much easier to visualize. The $H_0(\Sigma)$ counts the number of 
connected components of $\Sigma$. Let us, for simplicity, assume that $\Sigma$ is connected then $H_0(\Sigma) = \mathbb{R}$. The top homology $H_d(\Sigma)$ represents  $d$-dimensional cells with no boundary.  In absence of boundary, $\Sigma$ itself  is  the generator of $H_d(\Sigma) = \mathbb{R}$, but in presence of boundary $\p \Sigma \neq \emptyset$ and 
$H_d(\Sigma)=0$.

 For connected $d$-dimensional  $\Sigma$ with  boundary 
\be
H^0(\Sigma) =H_0 (\Sigma) = \mathbb{R},\;\;  H^d(\Sigma) = H_d(\Sigma) = 0
\ee
The difference in dimensions between $H^0(\Sigma)$ and $H^1(\Sigma)$ not only exclude the Poincare paring, but   any  non-degenerate paring (\ref{paring_candidate}) is excluded as well.

Our simple demonstration leads to the  conclusion: Cohomology groups  $H^\ast(\Sigma)$ are well defined in a presence of boundary, but the naive generalization  $\mathcal{M}^{inv} = H^p(\Sigma)\oplus H^{d-p}(\Sigma)$ fails, since $H^p(\Sigma)$ and $H^{d-p}(\Sigma)$ in general  have different dimensions,  so there is no non-degenerate paring. 

\section{Cohomology for manifold with boundary}\label{cohomology_review}
The invariant phase space for BF theory  from section \ref{BF_theory} has  description in terms of the de Rham cohomology. Since in presence of boundary de Rham cohomology fail to describe the invariant phase space, we can check the  algebraic topology for possible generalization.   The algebraic topology textbook \cite{bott2013differential} provides us with several immediate generalizations of de Rham cohomology that reduce to the (\ref{BF_inv_no_bound})  in absence of boundary.

\subsection{Cohomology with compact support}\label{compact_support}
 Compact support  $p$-form $A \in \Omega^{p}_c (\Sigma)$   is a $p$-form on $\Sigma$ which is  zero outside a compact set  $ C \subset \Sigma$.  The de Rham differential preserves this property, so we can define the corresponding cohomology, denoted as  $H_c^{p}(\Sigma)$. There is non-degenerate  Poincare paring 
\be\label{Poincare_comp}
H_c^p(\Sigma) \times H^{d-p} (\Sigma) \to \mathbb{R}: (A,B) \mapsto \int_{\Sigma} A \wedge B.
\ee
which provides us with the  phase space candidate
\be\label{BF_comp}
\mathcal{M}^{comp}_\Sigma =H_c^p(\Sigma)\oplus H^{p}_c(\Sigma)^\ast =  H_c^p(\Sigma)\oplus H^{d-p}(\Sigma)
\ee
 For connected $d$-dimensional  $\Sigma$ with  boundary 
\be
H^0(\Sigma)  = \mathbb{R},\;\;  H^d_c(\Sigma) =\mathbb{R},
\ee
so  the  paring  (\ref{Poincare_comp}) is non-degenerate. 
\subsection {Relative cohomology} 
For   $C \subset \Sigma$ we can define the relative $p$-chains as elements of 
 \be
 C_p (\Sigma, C) = C_p (\Sigma) / C_p(C).
 \ee
 The boundary operator $\p: C_p (\Sigma) \to C_{p-1}(\Sigma) $ naturally descends to quotient $ C_p (\Sigma, C) $, so we can define the corresponding homology $H_p(\Sigma,C)$, known as the {\it relative homology}.  The relative $p$-cycle $c$ is a $p$-chain, that can be anchored on $\p \Sigma$ i.e
 \be
c \in  Z_p (\Sigma, C)\; \Leftrightarrow \; \p c  \in C_{p-1}(\p \Sigma).
 \ee
The relative cohomology are part of the  Lefschetz paring  
\be\label{lefs}
H^p(\Sigma) \times H^{d-p} (\Sigma, \p \Sigma) \to \mathbb{R},
\ee
which we can use to define   the symplectic space
\be\label{BF_relative}
\mathcal{M}_\Sigma^{rel} =H^p(\Sigma) \oplus H^p(\Sigma)^\ast= H^p(\Sigma)\oplus H^{d-p}(\Sigma, \p \Sigma).
\ee
 For connected $d$-dimensional  $\Sigma$ with  boundary 
\be
H^0(\Sigma)  = \mathbb{R},\;\;  H^d(\Sigma, \p \Sigma) =\mathbb{R},
\ee
so  the paring (\ref{lefs}) is non-degenerate. Author of \cite{wu1991topological} used relative cohomology to describe  the phase space of BF theory.

\subsection  {Mapping cone cohomology} \label{mapping_cone}
 Let $f: S \to \Sigma$ be a smooth map, then we can define mapping cone differential forms 
\be
\Omega^p(f)  = \Omega^p (\Sigma) \oplus \Omega^{p-1}(S)
\ee
and differential 
\be
d_f :\Omega^p( f)  \to \Omega^{p+1}( f):\;  (B, b) \mapsto (dB, f^\ast B-db),\;\; d_f^2=0. 
\ee
The cohomology of  $(\Omega^\ast( f), d_f)$ are known as the {\it mapping cone} de Rham cohomology and denoted as $H^\ast (f)$. We can use 
and embedding map $i_{\p \Sigma}:\p \Sigma \hookrightarrow \Sigma$  to define the non-degenerate Poincare pairing 
\be\label{cone_pairing}
 H^p(\Sigma) \times H^{d-p} ( i_{\p \Sigma} ) \to \mathbb{R}: (A, (B,b)) \mapsto \int_{\Sigma} A \wedge B + \int_{\p \Sigma} A \wedge b
\ee
and the symplectic manifold
\be\label{BF_cone}
\mathcal{M}_\Sigma^{cone} = H^p(\Sigma) \oplus H^p(\Sigma)^\ast = H^p(\Sigma)\oplus H^{d-p}( i_{\p \Sigma}).
\ee
The mapping cone differential forms are very similar to  edge modes.  Authors of \cite{Mathieu:2019lgi} used mapping cone cohomology to analyze the invariant phase space of YM theory. 
\subsection{Cohomology relations} 
The three types of cohomology are related. The relative cohomology $H^\ast(\Sigma, C)$ are the same as compact support cohomology when $C$ is the boundary of $\Sigma$
\be
H^\ast(\Sigma, \p \Sigma)= H_c^\ast(\Sigma).
\ee 
The mapping  cone cohomology $H^\ast(f) $ are identical to the relative cohomology for $f$ being an embedding map $i_C : C \hookrightarrow \Sigma$
\be
H^\ast( i_C) = H^\ast (\Sigma, C).
\ee
Thus we conclude that   three types of cohomology are the same 
\be\label{cohomol_same}
H^\ast_c (\Sigma) = H^\ast (\Sigma, \p \Sigma) = H^\ast (i_{\p \Sigma}). 
\ee

\subsection{BF theory duality}
In case of the surface $\Sigma$ without boundary the invariant phase space (\ref{BF_inv_no_bound})  is invariant under the  $p \to d-p$ transformation. Neither of possible generalizations of the invariant phase space (\ref{BF_comp}), (\ref{BF_relative}) and (\ref{BF_cone}) is invariant
\be
\mathcal{M}^{(p) c}_\Sigma = H_c^p(\Sigma)\oplus H^{d-p}(\Sigma) \neq \mathcal{M}^{(d-p) c}_\Sigma = H_c^{d-p}(\Sigma)\oplus H^{p}(\Sigma). 
\ee
 For connected $d$-dimensional  $\Sigma$ with  boundary 
\beq
H^0 (\Sigma) = H^d_c(\Sigma) = \mathbb{R},\;\;\; H^0_c (\Sigma) = H^d(\Sigma)=0, \\
\mathcal{M}^{(0) c}_\Sigma = H_c^0(\Sigma)\oplus H^{d}(\Sigma)  = \mathbb{R}\oplus \mathbb{R}\neq \mathcal{M}^{(d) c}_\Sigma = H_c^{d}(\Sigma)\oplus H^{0}(\Sigma)  = 0.
\eeq

\section{Boundary terms and parings}\label{bound_pairing}
In section \ref{cohomology_review} we provided a brief review of the de Rham cohomology generalizations for the manifold with boundary. Each of discussed generalizations is connected to the generalization of the Poincare paring. The Poincare paring (\ref{BF_form_no_bound}) on  finite-dimensional  de Rham cohomology can be extended to the paring on the infinite-dimensional space of differential forms (\ref{bare_sympl_BF}).
In this section we propose the extension of the generalized Poincare pairings (\ref{Poincare_comp}) and (\ref{cone_pairing}) to the differential forms. Such extension allows us to  describe the phase space for the BF theory in the presence of boundary.

We use the triangulation of the manifolds to turn the infinite-dimensional spaces of differential forms into finite-dimensional. The existence of non-degenerate paring between two spaces require equality of the dimensions of the two spaces, what can be used to conjecture the structure of the paring. Let us note that the  BF theory is topological, so  the triangulated version of the theory to describe the invariant phase space identical to the continuous version. For simplicity of our argument we will focus on the $d=1$ theory, with only two possible topologies: the circle $S^1$ with no boundary and the interval $I$  with 
boundary being the pair of points.

\subsection{Triangulated  phase space}
We can triangulate the   $S^1$ by a  graph with $V$ vertices and $E$ edges. 
The spaces of simplexes of dimension $0$  and $1$ are 
\be
C_0(\Sigma)  = \mathbb{R}^V,\;\; C_1(\Sigma)  =  \mathbb{R}^E.
\ee
The forms on the triangulation are functions on simplexes 
\be
C^0(\Sigma) = Hom (C_0 (\Sigma), \mathbb{R} ) = C_0(\Sigma) =  \mathbb{R}^V,\;\;  C^1(\Sigma) = Hom (C_1 (\Sigma), \mathbb{R} ) = C_1(\Sigma) =  \mathbb{R}^E.
\ee
The triangulation of  $S^1$ is such that  $E=V$, so the spaces of 0-forms and 1-forms are identical
\be
C^0(\Sigma) = \mathbb{R}^V \simeq \mathbb{R}^E = C^1(\Sigma)
\ee
what we can equivalently reformulate as discrete version of Poincare duality
\be
\label{dis_Poincare}
C^0(\Sigma)^\ast  =  C^1(\Sigma),\;\;\; C^0(\Sigma)^\ast =C^0(\Sigma).
\ee
In case of an  interval  $\Sigma  = I = [0,1]$ with   boundary $\p I = \{0\}\sqcup\{1\}$
the triangulation has
\be
C_0(\Sigma) =  \mathbb{R}^V,\;\;  C_1(\Sigma) =  \mathbb{R}^E,
\ee
but  $V-E=1$, so there is no canonical isomorphism between $C_0$ and $C_1$.  Similar to the cohomology we can try to modify the (\ref{dis_Poincare}) by considering the compact support 
forms or using the mapping cone construction.

\subsection{Compactly supported forms}
Relative forms in our discrete model are forms that vanish on the boundary 
\beq
C^0 (\Sigma, \p \Sigma) = C^0 (\Sigma)/C^0 (\p\Sigma)  =  \mathbb{R}^V/\mathbb{R}^2 = \mathbb{R}^{V-2} \\
C^1 (\Sigma, \p \Sigma) = C^1 (\Sigma)/C^1 (\p\Sigma)  =  \mathbb{R}^E,
\eeq
so we can define embeddings 
\beq
C^0 (\Sigma, \p \Sigma) ^\ast  =\mathbb{R}^{V-2} \hookrightarrow   \mathbb{R}^{V-1} = \mathbb{R}^E =C^1(\Sigma),\\
C^1 (\Sigma, \p \Sigma) ^\ast=\mathbb{R}^E  \hookrightarrow   \mathbb{R}^{E+1} =  C^0(\Sigma).
\eeq
The  continuous version can be conjectured being
\be
\Omega^p(\Sigma, \p \Sigma)^\ast \sim  \Omega^{d-p}(\Sigma),
\ee
while the pairing is
\be\label{copact_form_par}
 \Omega^p (\Sigma, \p \Sigma) \times  \Omega^{d-p}(\Sigma)\to \mathbb{R}: (A;B) \mapsto \int_\Sigma A \wedge B.
\ee

\subsection{Mapping cone forms} 
Mapping cone forms for the triangulation of an interval $\Sigma$ with boundary $\p \Sigma$ 
\beq
 C^0 (i_{\p \Sigma})= C^0(\Sigma)\oplus C^{-1}(\p \Sigma) =  \mathbb{R}^V\\
  C^1 (i_{\p \Sigma})= C^1(\Sigma)\oplus C^0(\p \Sigma) =  \mathbb{R}^E\oplus \mathbb{R}^2 = \mathbb{R}^{E+2},
\eeq
so  can define  embeddings 
\beq
C^1(\Sigma)^\ast  =\mathbb{R}^E \hookrightarrow \mathbb{R}^{E+1}  =  \mathbb{R}^V =C^0(\Sigma),\\
 C^0(\Sigma)^\ast=\mathbb{R}^V = \mathbb{R}^{E+1} \hookrightarrow   \mathbb{R}^{E+2}= C^1 (i_{\p \Sigma}).
\eeq
In case or dimensions $d$ higher then $1$ and arbitrary $p$ 
\be
C^p (\Sigma)^\ast \hookrightarrow C^{d-p}(\Sigma) \oplus C^{d-p-1}(\p \Sigma)=C^{d-p} (i_{\p \Sigma}).
\ee
The  continuous version is conjectured to be 
\be
\Omega^p(\Sigma)^\ast \sim \Omega^{d-p}(\Sigma)\oplus \Omega^{d-p-1}(\p \Sigma) = \Omega^{d-p}(i_{\p \Sigma}),
\ee
while the pairing is 
\be\label{map_cone_paring}
 \Omega^p (\Sigma) \times  \Omega^{d-p}(i_{\p \Sigma})\to \mathbb{R}: (A;B,b) \mapsto \int_\Sigma A \wedge B +\int_{\p \Sigma} A \wedge b.
\ee

\subsection{Phase spaces}\label{phase_spaces}
Using the two pairings (\ref{copact_form_par}) and (\ref{map_cone_paring}) we can construct two linear  symplectic spaces for the  surface $\Sigma$ with  boundary:
\begin{itemize}
\item {\it Compact field phase space}
\be\label{M_BF_comp}
\mathcal{M}_\Sigma^{comp} = \Omega^p(\Sigma, \p \Sigma) \oplus  \Omega^{d-p}(\Sigma),
\ee
 with  symplectic form 
\be
\omega^{comp}_\Sigma=- \int_\Sigma \delta A\wedge \delta B.  
\ee
\item {\it Edge mode phase space}
\be\label{M_BF_bound}
\mathcal{M}_\Sigma^{edge}= \Omega^p(\Sigma) \oplus \Omega^{d-p}(\Sigma)\oplus \Omega^{d-p-1}(\p \Sigma).
\ee
endowed with  symplectic form 
\be
\omega^{edge}_\Sigma=- \int_\Sigma \delta A\wedge \delta B - \int_{\p \Sigma} \delta A \wedge \delta b. 
\ee
The field $b \in \Omega^{d-p-1}(\p \Sigma)$ is commonly referred as the {\it edge mode}.
\end{itemize}
The two phase spaces  $\mathcal{M}_\Sigma^{comp}$  and $\mathcal{M}_\Sigma^{edge}$  are related by the symplectic reduction
\be
\mathcal{M}^{comp}_\Sigma =\mathcal{M}^{edge}_\Sigma // S^{\p \Sigma},
\ee
where  $S^{\p \Sigma}$ is the {\it surface symmetry},  describing the   redefinition of $b$
\be
\delta b  = \sigma,\;\; \sigma \in \Omega^{d-p-1}(\p \Sigma).
\ee
The corresponding Hamiltonian is
\be
H_\sigma = (-1)^{p(d-p-1)} \int_{\p \Sigma} \sigma \wedge A,
\ee
while the   moment map is
\be
\mu_S  = (-1)^{p(d-p-1)} i^\ast_{\p \Sigma}A  =  (-1)^{p(d-p-1)}A|_{\p \Sigma}\in  \mathfrak{s}^{\p \Sigma\ast} \simeq\Omega^p (\p \Sigma).
\ee
The zero locus of moment map  is
\be
\mu_S^{-1}(0) = \Omega^p (\Sigma, \p \Sigma) \oplus \Omega^{d-p} (\Sigma)\oplus \Omega^{d-p-1}(\p \Sigma),
\ee
while we can use the surface symmetry to set $b=0$,  so that 
\be
\mathcal{M}^{red}_\Sigma =\mathcal{M}^{edge}_\Sigma // S^{\p \Sigma} =   \mu_S^{-1}(0)/S^{\p\Sigma} =  \Omega^p (\Sigma, \p \Sigma) \oplus \Omega^{d-p} (\Sigma) = \mathcal{M}^{comp}_\Sigma.
\ee
The symplectic form  
\be
\omega^{red}_\Sigma  =  \pi^\ast i_\mu^\ast \omega^{edge}_{\Sigma}=- \int_\Sigma \delta A\wedge \delta B = \omega_\Sigma^{comp}.  
\ee

\subsection{Gauge symmetries}
The algebra of gauge symmetries in BF theory  can be identified with differential forms on $\Sigma$
\be
\mathfrak{g}^\Sigma = Lie(G^\Sigma) = \Omega^{p-1}(\Sigma),\;\;\; \mathfrak{k}^\Sigma  = Lie (K^\Sigma)= \Omega^{d-p-1}(\Sigma).
\ee
The symplectic reduction theorem requires the  moment map, defined from  canonical pairing  between $\mathfrak{g}$ and $\mathfrak{g}^\ast$. For the surface without boundary   the Poincare duality allowed us to identify 
\be
\mathfrak{g}^{\Sigma\ast} = \Omega^{p-1}(\Sigma)^\ast  = \Omega^{d-p+1}(\Sigma),\;\;\; \mathfrak{k}^{\Sigma\ast} = \Omega^{d-p-1}(\Sigma)^\ast  = \Omega^{p+1}(\Sigma),
\ee
while in presence of boundary we need to use the modified pairing (\ref{map_cone_paring}) to describe the dual algebras 
\beq\label{gauge_sym_dual}
\mathfrak{g}^{\Sigma\ast} = \Omega^{p-1}(\Sigma)^\ast  = \Omega^{d-p+1}(\Sigma)\oplus \Omega^{d-p}(\p \Sigma),\\
 \mathfrak{k}^{\Sigma\ast} = \Omega^{d-p-1}(\Sigma)^\ast  = \Omega^{p+1}(\Sigma)\oplus \Omega^{p}(\p \Sigma).
\eeq

The moment map for the infinitesimal action of gauge symmetry $G^\Sigma$ on hypersurface $\Sigma$ with boundary has two components: one with values in bulk forms   $\Omega^{d-p+1} (\Sigma)$, while the other with values  in boundary forms $\Omega^{d-p} (\p \Sigma)$.  

The bulk form part of the moment map for $G^\Sigma$ is the moment map for
\be
G_c^\Sigma = \{ g\in G^\Sigma\;|\; g|_{\p \Sigma}=\hbox{I}\},
\ee
a subgroup of all gauge transformations $G^\Sigma$, trivial at the boundary $\p \Sigma$ of $\Sigma$. Such subgroup often referred in a literature as a compactly supported gauge transformations.  
The Lie algebra   of $G^\Sigma_c$  is $(p-1)$-forms on $\Sigma$ that vanish on the boundary i.e.
\be
\mathfrak{g}_c^\Sigma = \Omega^{p-1}(\Sigma, \p \Sigma),
\ee
while the dual Lie algebra 
\be
\mathfrak{g}_c^{\Sigma\ast} \simeq \Omega^{d-p+1} (\Sigma) 
\ee
has natural embedding 
\be
i_c : \mathfrak{g}_c^{\Sigma\ast} \hookrightarrow\mathfrak{g}^{\Sigma\ast}:\; f \mapsto (f,0).
\ee
The moment map for $G^\Sigma_c$-action is a pullback of the $G^\Sigma$-moment map  
\be
\mu_{G_c} = i_c^\ast \mu_G,\;\; \mu_{G_c} = i_c^\ast (f,a) = f.
\ee
The   subspace $\mu_{G_c}=0$ is the the space of solutions to the  Gauss law constraint. 

\subsection{Asymptotic symmetries}
We can perform the symplectic reduction over $G^\Sigma$-action  in two steps. The  $G_c^\Sigma$ reduction first  
\be
\mathcal{M}_{\Sigma}^{inv,c} = \mathcal{M}_\Sigma//G_c^\Sigma = \frac{\mu_{G_c}^{-1}(0)}{G_c^\Sigma}
\ee
and  the second reduction   under the  action of  stabilizer group of zero  
\be
 G^\Sigma/G^\Sigma_c = G^{\p \Sigma} : \mathfrak{g}^{\Sigma\ast}/\mathfrak{g}_c^{\Sigma\ast} \to \mathfrak{g}^{\Sigma\ast}/\mathfrak{g}_c^{\Sigma\ast}.
\ee
 According to the reduction in stages   the phase space $\mathcal{M}_{\Sigma}^{inv,c}$ carries the quotient group action, which can be naturally identified with the asymptotic symmetry group. 
The commonly used definition of the {\it asymptotic symmetry group} (ASG) is 
\be
ASG = \frac{\hbox{Allowed transformations}}{\hbox{Trivial transformations}},
\ee 
where {\it allowed transformations} defined as (infinitesimal gauge) transformations $\e$ with finite values of a certain  boundary charge $Q[\e]$, while the {\it trivial transformations} are the ones with zero values of $Q[\e]$.
In our notations we can construct  the boundary charge  $Q[\e]$  from the moment map $\mu_{\p \Sigma}$ of the $G^{\p \Sigma}$-action
\be
Q[\e]= \<\e,i_\partial^\ast \mu_{\p \Sigma}\> = \int_{\p \Sigma} i_{\p \Sigma}^\ast\e \wedge a,
\ee
for the gauge transformation parameter $\e \in \mathfrak{g}^{\Sigma}$ and  embedding 
\be
i_\partial :\mathfrak{g}^{\Sigma\ast}/ \mathfrak{g}_c^{\Sigma\ast} \hookrightarrow\mathfrak{g}^{\Sigma\ast}:\; a \mapsto (0,a).
\ee

\section{Symplectic reduction in presence of boundary}
The invariant phase space  of the BF theory (\ref{BF}) can be defined as the  symplectic reduction of the edge-mode extended  phase  $\mathcal{M}^{edge}_\Sigma$ over the action of 
gauge symmetries $G^\Sigma \times K^\Sigma $  and surface symmetry $S^{\p \Sigma}$
\be\label{BF_inv_def}
\mathcal{M}^{inv}_\Sigma =  \mathcal{M}^{edge}_\Sigma // (G^\Sigma\times K^\Sigma \times S^{\p \Sigma}).
\ee
In this section we will carefully describe the symmetries and describe the symplectic reduction.  

\subsection{Symmetries }\label{BF_symm_central}
Let us identify the symplectic symmetries of our system. There two types of gauge symmetries  $ G^\Sigma$ for field $A$ and $K^\Sigma$ for $B$-field with infenitesimal versions
\be
\delta A =  d\e,\;\; \delta B = d\lambda,\;\;  \e \in \mathfrak{g}^\Sigma = \Omega^{p-1}(\Sigma),\;\; \lambda\in \mathfrak{k}^\Sigma=\Omega^{d-p-1}(\Sigma). 
\ee
The corresponding Hamiltonians
\be\label{Ham_gauge_boundary}
\begin{split}
H_\e &= (-1)^{p+1}\int_{\Sigma} \e \wedge dB -\int_{\p \Sigma} \e \wedge (B+(-1)^{p} db),\\
 H_\lambda  &= (-1)^{(p+1)(d-p)}  \int_{\Sigma} \lambda  \wedge d A + (-1)^{p(d-p)} \int_{\p \Sigma} \lambda  \wedge  A .
\end{split}
\ee
The  surface symmetry $S^{\p \Sigma}$ is the edge mode redefinition 
\be
\delta b  = \sigma,\;\; \sigma \in \Omega^{d-p-1}(\p \Sigma),
\ee
with Hamiltonian
\be\label{Ham_surf_bound}
H_\sigma =(-1)^{p(d-p-1)}  \int_{\p \Sigma} \sigma \wedge A.
\ee
From  (\ref{gauge_sym_dual})  we can deduce  the moment maps 
\beq\label{moment_bf_bound}
\mu_G = ((-1)^{p+1} dB, - i^\ast_{\p \Sigma}B +(-1)^{p+1}db) \in \mathfrak{g}^{\Sigma\ast} = \Omega^{d-p+1}(\Sigma)\oplus \Omega^{d-p}(\p \Sigma), \\
\mu_K = ((-1)^{(p+1)(d-p)}  dA, (-1)^{p(d-p)}  i^\ast_{\p \Sigma}A) \in \mathfrak{k}^{\Sigma\ast} = \Omega^{p+1}(\Sigma)\oplus \Omega^{p}(\p \Sigma), \\
\mu_S =(-1)^{p(d-p-1)}  i^\ast_{\p \Sigma}A \in \mathfrak{s}^{\p\Sigma\ast} = \Omega^{p}(\p \Sigma). 
\eeq

All groups $G^\Sigma, H^\Sigma,S^{\p \Sigma} $ are abelian so their   Lie algebras have trivial brackets.  The Poisson algebra of (\ref{Ham_gauge_boundary}), (\ref{Ham_surf_bound})  is centrally extended 
\be\label{algebra_gauge_bound}
\begin{split}
\{H_\e, H_\lambda\} &=(-1)^{p(d-p)} \int_{\p \Sigma} \lambda \wedge d\e,\\ 
\{H_\e, H_\sigma\} &=(-1)^{p(d-p-1)}\int_{\p \Sigma} \e \wedge d\sigma,\\
 \{H_\lambda, H_\sigma\} &=0,
\end{split}
\ee
so the corresponding moment maps are not equivariant.

 Let us point out that the central extension in (\ref{algebra_gauge_bound}) is a generic feature of the phase space with edge modes. To describe the  phase space for  all $A$-field configurations, including the ones not vanishing on the boundary, i.e. $A \in \Omega^p (\Sigma)$ we need to enlarge the "dual momenta" $B \in \Omega^{d-p}(\Sigma)$ by an edge mode $b \in \Omega^{d-p}(\p \Sigma)$. We can freely redefine the edge mode $b$ by the surface symmetry action.  The moment map for the surface symmetry  is  the boundary value $i^\ast_{\p \Sigma}A$ of $A$. The boundary value is shifted by the boundary gauge transformations, hence the surface symmetry moment map is not equivariant. The non-equivariance of the moment map   is equivalent to the nontrivial central extension for the phase space realization of symmetries.

\subsection{Reduction in stages}\label{bf_red_stages}
The symplectic space  (\ref{BF_inv_def}) is defined as symplectic reduction with respect to the non-equivariant action of gauge and surface symmetries. The non-equivariant reduction theorem in section \ref{non_equiv_reduct}
provides us with construction to  represent this reduction as a quotient space. However, we are not going to use this method, since one of the goals of this note is to compare our results for BF theory with
similar results from the literature.  We can decompose the group of all symmetries into parts and perform the reduction in stages. Not only we can mach the results from literature but also we can avoid usage of the non-equivariant reduction. Let us outline the the two decompositions below, while leaving the details of  reductions in following sections.
\begin{itemize}

\item {\it Compact support description}:
We can perform a symplectic reduction over $S^{\p \Sigma}$ first, so  the invariant phase space (\ref{BF_inv_def}) becomes
\be
\mathcal{M}^{inv}_{\Sigma}  =  \left[\mathcal{M}^{edge}_\Sigma // S^{\p \Sigma}\right] //(G^\Sigma\times K^\Sigma )^a_0.
\ee 
The advantage of this decomposition is that the first reduction is easy to perform as it is a familiar compact  support version of the phase space 
\be
\mathcal{M}^{comp}_\Sigma =\mathcal{M}^{edge}_\Sigma // S^{\p \Sigma} = \Omega^p(\Sigma, \p \Sigma)\oplus \Omega^{d-p}(\Sigma).
\ee
The reduction over $(G^\Sigma\times K^\Sigma )^a_0$ can be further decomposed into the compact support gauge transformations and the boundary support one, what makes it very similar to the 
asymptotic symmetry approach towards the invariant phase space construction.

\item {\it Edge modes}:  We can also observe that both central terms in  (\ref{algebra_gauge_bound})  have similar structure 
so we can cancel them if we mix $K^\Sigma$- and $S^{\p \Sigma}$- symmetries  in a specific way.  Let denote the modified symmetry by $\tilde{K}^\Sigma$, while the reduction in stages gives us another description of the invariant phase space
\be
\mathcal{M}^{inv}_{\Sigma}  =  \left[\mathcal{M}^{edge}_\Sigma  //(G^\Sigma\times \tilde{K}^\Sigma )\right] //(S^{\p \Sigma})^a_0
\ee 
The  phase space in square brackets can be naturally identified with the edge mode construction of phase space. 

\end{itemize}

\subsection{Invariant phase space with edge modes}
Let us define the new symmetry  $\tilde{K}^{\Sigma}$ as a  combination of  $K^\Sigma$ and $S^{\p\Sigma}$ with the infenitesimal action 
\be\label{modified_gauge_BF}
 \delta_g (B,b) =(d\lambda, (-1)^{p+1}i_{\p \Sigma}^\ast\lambda +d \xi) = d_{i_{\p \Sigma}}(\lambda, \xi),\;\;  \lambda\in \Omega^{d-p-1}(\Sigma),\;\; \xi  \in \Omega^{d-p-2}(\p \Sigma),
\ee
so it is identical to the de Rham differential for the mapping cone from section \ref{mapping_cone} up to a field redefinition by a multiplicative factor $(-1)^p$.
The modified Hamiltonian
\be
\tilde{H}_{ \lambda} =  (-1)^{(p+1)(d-p)}  \int_{\Sigma} \lambda  \wedge  dA    +(-1)^{(p+1)(d-p-1)}  \int_{\p \Sigma}\xi\wedge d A 
\ee
has trivial Poisson bracket with   $H_\e$
\be
\{ \tilde{H}_\lambda, H_\e\} =0,
\ee
We manage   to remove the central extension term in (\ref{algebra_gauge_bound}) by the symmetry modification. 
The  symplectic reduction
\be
\mathcal{M}^{edge,inv}_\Sigma = \mathcal{M}^{edge}_\Sigma // (G^\Sigma \times \tilde{K}^\Sigma)
\ee
is an equivariant moment map reduction.  
The moment maps for the  group actions 
\beq
\mu_G= ((-1)^{p+1} dB, - i^\ast_{\p \Sigma}B +(-1)^{p+1}db) \in  \Omega^{d-p+1}(\Sigma) \oplus \Omega^{d-p}(\p \Sigma) \\
\mu_{\tilde{K}}=  ( (-1)^{(p+1)(d-p)} dA, (-1)^{(p+1)(d-p-1)} d i^\ast_{\p \Sigma} A)\in \Omega^{p+1}(\Sigma)\oplus\Omega^{p+1}(\p \Sigma).
\eeq
  The zero locus of the moment map 
\be
\mu_G^{-1}(0) \cap \mu_{\tilde{K}}^{-1}(0) = Z^p(\Sigma) \times Z^{d-p}(i_{\p \Sigma}),
\ee
with  $Z^{d-p}(i_{\p \Sigma} )$ being the space of closed $(d-p)$-forms in  the   mapping cone de Rham complex  form section \ref{mapping_cone}.
The invariant phase space  phase space after reduction 
\be\label{BF_edge_inv}
\mathcal{M}^{edge,inv}_\Sigma = \frac{\mu_G^{-1}(0) \cap \mu_{\tilde{K}}^{-1}(0)}{ G^\Sigma \times \tilde{K}^\Sigma} =  \frac{Z^{p}(\Sigma)\oplus Z^{d-p}(i_{\p \Sigma})}{ d\Omega^{p-1}(\Sigma)\oplus d_{i_{\p \Sigma}}\Omega^{d-p-1}(i_{\p \Sigma})} = 
H^{p}(\Sigma) \oplus H^{d-p}(i_{\p \Sigma})
\ee
 is identical to the mapping cone generalization of the invariant phase space (\ref{BF_comp}).   
The symplectic form 
\be\label{canonical_BF_inv}
\omega^{edge,inv}_\Sigma  = s^\ast i_\mu^\ast \omega^{edge}_\Sigma= -\int_\Sigma \delta A\wedge \delta B -\int_{\p\Sigma} \delta A\wedge \delta b  ,\;\; A \in H^{p}(\Sigma),\;\; (B, b) \in H^{d-p}(i_ {\p \Sigma})
\ee
is canonical symplectic from for the pairing (\ref{cone_pairing}) and is similar to the results in a literature \cite{Donnelly:2016auv} for Maxwell theory.

The remaining part of the surface symmetry $(S^{\p \Sigma})^a_0$ is defined as the stabilizer of $0$ under the the affine action of the $S^{\p \Sigma}$. The action of $S^{\p \Sigma}$ on the moment maps
\beq
\delta_S \mu_G=\delta_S  ((-1)^{p+1} dB, - i^\ast_{\p \Sigma}B +(-1)^{p+1}db) = (0, (-1)^{p+1}d\sigma) \\
\delta_S \mu_{\tilde{K}}= \delta_S  ( (-1)^{(p+1)(d-p)} dA, (-1)^{(p+1)(d-p-1)} d i^\ast_{\p \Sigma} A) = (0,0).
\eeq
The subgroup of $S^{\p \Sigma}$ that preserves the zero locus of moment map is $Z^{d-p-1}(\p \Sigma)$, the closed forms on the boundary. The exact forms on the boundary are already included in 
(\ref{modified_gauge_BF}) so we left with the  cohomology $H^{d-p-1}(\p \Sigma)$. The invariant phase space  becomes 
\be\label{inv_edge}
\mathcal{M}^{inv}_{\Sigma}   = \mathcal{M}^{edge,inv}_\Sigma  // H^{d-p-1}(\p \Sigma).
\ee
The moment map of this action 
\be
\mu_{S^{\p \Sigma}}: \mathcal{M}^{edge, inv}_\Sigma \to  H^{p}(\p \Sigma): (A, B, b) \mapsto (-1)^{p(d-p-1)}  i^\ast_{\p \Sigma}A. 
\ee
The zero locus of  the moment map 
\be
\mu_{S^{\p \Sigma}}^{-1}(0) = \ker\left[ i_{\p \Sigma}^\ast:  H^{p}(\Sigma) \to H^p (\p \Sigma)\right] \oplus H^{d-p}(i_{\p \Sigma}),
\ee
while the  reduced phase space is 
\be\label{reduced_pahse_edge_coh}
\begin{split}
\mathcal{M}^{inv}_\Sigma &= \frac{\mu_{ S^{\p \Sigma}}^{-1}(0) }{ (S^{\p \Sigma})^a_0} \\
&= \ker\left[ i_{\p \Sigma}^\ast:  H^{p}(\Sigma) \to H^p (\p \Sigma)\right] \oplus \frac{H^{d-p} (i_{\p\Sigma}) }{ \hbox{Im} \left[i: H^{d-p-1}(\p \Sigma) \to H^{d-p}(i_{\p \Sigma})\right] },
\end{split}
\ee
where $i: H^{d-p-1}(\p \Sigma) \to H^{d-p}(i_{\p \Sigma})$  is an embedding map for short exact sequence associated to the mapping cone cohomology. The mapping cone forms can be organized into 
a short exact sequence 
\be
\xymatrix@1{
0 \ar[r] & \Omega^{d-p-1}(\p \Sigma) \ar[r]^i & \Omega^{d-p}(i_{\p \Sigma}) \ar[r]^\pi & \Omega^{d-p}(\Sigma) \ar[r]&0 \\
& b\;\;\; \ar[r]& \;\;\;(0,b)\;\;&&\\
&&\;\;\;(B,b)\;\;\ar[r]&\;\;\;B&.
}
\ee
Given a short exact sequence we can construct long exact sequence of cohomology
\be\label{long_exact_edge}
\xymatrix@1{
...\ar[r] &H^{d-p-1}(\Sigma)\ar[r]^{i_{\p \Sigma}^\ast} & H^{d-p-1}(\p \Sigma )\ar[r]^i &H^{d-p}(i_{\p \Sigma})\ar[r]^\pi &H^{d-p}(\Sigma)\ar[r]^{i_{\p \Sigma}^\ast} &.... 
}
\ee
which defines the  map $i: H^{d-p-1}(\p \Sigma) \to H^{d-p}(i_{\p \Sigma})$.

We can use the long exact sequence (\ref{long_exact_edge}) to simplify our expression (\ref{reduced_pahse_edge_coh}) for the reduced phase space.  Since all cohomology are over real numbers the exact sequence splits and we can further rewrite 
\be
\frac{H^{d-p} (i_{\p\Sigma}) }{ \hbox{Im} (i)} = \hbox{Im} (\pi) = \ker \left[i_{\p \Sigma}^\ast: H^{d-p}(\Sigma) \to H^{d-p}(\p \Sigma) \right],
\ee
so that 
\be\label{BF_inv}
\mathcal{M}^{inv}_\Sigma =  \ker \left[ i_{\p \Sigma}^\ast: H^{p}(\Sigma) \to H^{p}(\p \Sigma) \right] \oplus  \ker \left[ i_{\p \Sigma}^\ast: H^{d-p}(\Sigma) \to H^{d-p}(\p \Sigma) \right].
\ee
Using a pair of long exact sequences (\ref{long_exact_edge}), arranged so that the vertical lines are generalized Poincare duality isomorphisms  
\be\notag
\xymatrix@1{
...\ar[r]&H^{p-1}(\Sigma)\ar[r]^{i_{\p \Sigma}^\ast}\ar[d]& H^{p-1}(\p \Sigma )\ar[r]^i \ar[d]&H^{p}(i_{\p \Sigma})\ar[r]^\pi \ar[d]&H^{p}(\Sigma)\ar[r]^{i_{\p \Sigma}^\ast}\ar[d]&H^{p}(\p \Sigma)\ar[r] \ar[d]&.... \\
...&\ar[l]H^{d-p+1}(i_{\p \Sigma})& \ar[l]^{i} H^{d-p}(\p \Sigma ) &\ar[l]^{i_{\p \Sigma}^\ast}H^{d-p}(\Sigma)& \ar[l]^\pi H^{d-p}(i_{\p \Sigma}) &\ar[l]^{i}H^{d-p-1}(\p \Sigma)&\ar[l]....
}
\ee
we can rewrite
\be
\mathcal{M}^{inv}_\Sigma =  \ker i^\ast_{\p \Sigma} \oplus \hbox{coker}\;i =  \ker i^\ast_{\p \Sigma} \oplus (\ker i^\ast )^\ast  = \ker i^\ast_{\p \Sigma} \oplus (\ker i^\ast_{\p \Sigma})^\ast,
\ee
so it assumes the form of the  linear symplectic space 
\be
\mathcal{M}^{inv}_\Sigma  = V\oplus V^\ast,\;\; V =  \ker i^\ast_{\p \Sigma}.
\ee

\subsection{Invariant phase space with  compact support} 
The action of  $(G^\Sigma\times K^\Sigma )$ on a  moment map (\ref{moment_bf_bound})  for the surface symmetry
\be
K^\Sigma\times G^\Sigma : \mu_{S}  =(-1)^{p(d-p-1)} i^\ast_{\p \Sigma} A \to (-1)^{p(d-p-1)} i^\ast_{\p \Sigma} A +(-1)^{p(d-p-1)} i^\ast_{\p \Sigma}d\e,
\ee
leads  to the stabilizer subgroup of the affine action of the form
\be
(K^\Sigma\times G^\Sigma )^a_0 = K^\Sigma\times G_c^\Sigma \times H^{p-1}(\p \Sigma).
\ee
The action of the $(K^\Sigma\times G^\Sigma )^a_0$ is equivariant since the only nontrivial Poisson bracket
\be
\{H_\e, H_\lambda\} = (-1)^{p(d-p)}  \int_{\p \Sigma} \lambda \wedge d\e =0
\ee
vanishes because  $i^\ast_{\p\Sigma}d\e=0$ follows  from the stabilizer  definition.

The compact support phase space 
\be
\mathcal{M}^{edge}_\Sigma // S^{\p \Sigma} = \mathcal{M}^{comp}_\Sigma  = \Omega^p(\Sigma, \p \Sigma) \oplus \Omega^{d-p}(\Sigma).
\ee
  is invariant under the compact support gauge transformations $G_c^\Sigma \times K^\Sigma$ with the corresponding moment maps 
\be
\begin{split}
\mu_{G_c}  &= (-1)^{p+1} dB \in \mathfrak{g}^{\Sigma \ast}_c  = \Omega^{d-p+1}(\Sigma),\\
 \mu_K  &= ((-1)^{(p+1)(d-p)} dA,0) \in \mathfrak{k}^{\Sigma \ast} = \Omega^{p+1}(\Sigma) \oplus \Omega^{p}(\p \Sigma).
 \end{split}
\ee 
The  reduced phase space 
\be\label{BF_comp_inv}
\mathcal{M}^{comp,inv}_\Sigma =\frac{ \mu^{-1}_{G_c}(0)\cap \mu_K^{-1}(0) }{ G_c^\Sigma \times K^\Sigma} =  H_c^p (\Sigma) \oplus H^{d-p}(\Sigma)
\ee
has natural description as a cohomology phase space (\ref{BF_comp}). The invariant phase space requires the final reduction 
\be\label{inv_asympt}
\mathcal{M}^{inv}_{\Sigma}   = \mathcal{M}^{comp,inv}_\Sigma // H^{p-1}(\p \Sigma).
\ee
We can describe using the $\mathcal{M}^{inv}_{\Sigma} $ using the  long exact sequence for relative cohomology and identification $H_c(\Sigma) = H(\Sigma, \p \Sigma)$.  
The short exact sequence associated to the relative forms
\be
\xymatrix@1{
0 \ar[r] & \Omega^{p}(\Sigma, \p\Sigma)   \ar[r]^{j^\ast} & \Omega^{p}(\Sigma) \ar[r]^{i^\ast_{\p\Sigma}} &\Omega^{p}(\p \Sigma) \ar[r]&0 \\
}
\ee
with $i_{\p\Sigma}^\ast$ being the restriction map of the differential forms on $\Sigma$ to $\p \Sigma$. The map $j^\ast:\Omega^{p}(\Sigma) \to \Omega^{p}(\Sigma, \p\Sigma) $ is a dual of the quotient map for chains
\be
j: C_p(\Sigma, \p\Sigma) = C_p(\Sigma)/C_p(\p\Sigma) \to C_p(\Sigma).
\ee
The corresponding  long exact sequence of cohomology takes the form
\be\label{long_relative}
\xymatrix@1{
...\ar[r] &H^{p-1}(\Sigma)\ar[r]^{i_{\p \Sigma}^\ast} & H^{p-1}(\p \Sigma )\ar[r]^i &H^{p}(\Sigma,\p \Sigma)\ar[r]^{j^\ast} &H^{p}(\Sigma)\ar[r]^{i_{\p \Sigma}^\ast}&H^{p}(\p \Sigma)\ar[r] &.... 
}
\ee
The  connecting homomorphism $i$ defines the  action of $H^{p-1}(\p \Sigma)$ on phase space  in (\ref{inv_asympt}). The moment map $\mu_i$ for this action can be described using the 
generalized Poincare duality
\be
(H^p(\p \Sigma))^\ast = H^{d-p}(\p\Sigma),\;\;\; (H^{p}(\Sigma,\p\Sigma))^\ast = H^{d-p}(\Sigma),
\ee
so that 
\be
i^\ast: (H^{p}(\Sigma,\p\Sigma))^\ast \to (H^p(\p \Sigma))^\ast : H^{d-p}(\Sigma) \to H^{d-p}(\p\Sigma)
\ee
and 
\be
\mu_i :  \mathcal{M}^{comp,inv}_\Sigma \to (H^{p-1}(\p \Sigma))^\ast: H^p (\Sigma, \p\Sigma) \oplus H^{d-p}(\Sigma) \to  H^{d-p}(\p\Sigma): (A,B )\mapsto i^\ast B.
\ee 
Using a pair of long exact sequence (\ref{long_relative}), arranged so that the vertical lines are generalized Poincare duality isomorphisms  
\be\notag
\xymatrix@1{
...\ar[r]&H^{p-1}(\Sigma)\ar[r]^{i_{\p \Sigma}^\ast}\ar[d]& H^{p-1}(\p \Sigma )\ar[r]^i \ar[d]^\simeq&H^{p}(\Sigma,{\p \Sigma})\ar[r]^{j^\ast} \ar[d]&H^{p}(\Sigma)\ar[r]^{i_{\p \Sigma}^\ast}\ar[d]&H^{p}(\p \Sigma)\ar[r] \ar[d]&.... \\
...&\ar[l]H^{d-p+1}(\Sigma,{\p \Sigma})& \ar[l]^{i} H^{d-p}(\p \Sigma ) &\ar[l]^{i_{\p \Sigma}^\ast}H^{d-p}(\Sigma)& \ar[l]^{j^\ast} H^{d-p}(\Sigma, {\p \Sigma}) &\ar[l]^{i}H^{d-p-1}(\p \Sigma)&\ar[l]....
}
\ee
we can argue that $i^\ast  = i_{\p\Sigma}^\ast$, i.e. it is a restriction map for the differential forms.

The quotient space description of the invariant phase space  (\ref{inv_asympt})
\be
\begin{split}
\mathcal{M}^{inv}_{\Sigma}  & = \mathcal{M}^{comp,inv}_\Sigma // H^{p-1}(\p \Sigma) = \frac{\mu_i^{-1}(0)}{\hbox{Im}(i)} = \frac{ \ker i^\ast \oplus H^p (\Sigma, \p\Sigma)}{\hbox{Im}(i)}\\
&=\ker i_{\p\Sigma}^\ast \oplus \frac{H^p (\Sigma, \p\Sigma)}{\hbox{Im}(i)}
\end{split}
\ee
In more explicit form the expression above is
\be
\mathcal{M}^{inv}_{\Sigma}  =\ker \left[ i_{\p \Sigma}^\ast :  H^{d-p}(\Sigma) \to H^{d-p} (\p \Sigma)\right]   \oplus \frac{H^p (\Sigma, \p\Sigma)}{\hbox{Im}\left[ i:  H^{p-1}( \p\Sigma) \to H^p (\Sigma, \p \Sigma)\right]}.
\ee
Using the long exact sequence (\ref{long_relative}) we can evaluate 
\be\notag
\begin{split}
\frac{H^p (\Sigma, \p\Sigma)}{\hbox{Im}\left[ i:  H^{p-1}( \p\Sigma) \to H^p (\Sigma, \p \Sigma)\right]} =& {\hbox{Im}\left[ j^\ast:  H^p (\Sigma, \p \Sigma) \to H^{p}( \Sigma)  \right]} \\
&=\ker \left[ i_{\p \Sigma}^\ast :  H^{p}(\Sigma) \to H^{p} (\p \Sigma)\right], 
\end{split}
\ee
so that 
\be
\mathcal{M}^{inv}_{\Sigma}   =\ker \left[ i_{\p \Sigma}^\ast :  H^{d-p}(\Sigma) \to H^{d-p} (\p \Sigma)\right] \oplus \ker \left[ i_{\p \Sigma}^\ast :  H^{p}(\Sigma) \to H^{p} (\p \Sigma)\right], 
\ee
which is identical to the phase space (\ref{BF_inv}), constructed using edge modes in previous section.

\subsection{Invariant phase space}\label{inv_phase_features}
We constructed the invariant phase space $\mathcal{M}^{inv}_{\Sigma}  $ for BF theory using two different decompositions of the symmetry group. In both cases 
we first performed the infinite-dimensional symplectic reduction to  get the finite-dimensional spaces $\mathcal{M}^{edge,inv}_\Sigma $ and $\mathcal{M}^{comp,inv}_\Sigma $. On second step we performed the finite-dimensional reductions for both spaces to construct the  $\mathcal{M}^{inv}_{\Sigma} $. We can describe  spaces   $\mathcal{M}^{edge,inv}_\Sigma $ and $\mathcal{M}^{comp,inv}_\Sigma $ as a   ``symplectic extension" for the invariant phase space by either the edge modes inclusion  or   asymptotical symmetries ``ungauging". 

The edge mode extension is self-explanatory, while the for the asymptotic symmetries ``ungauging" is the following procedure. By construction the invariant phase space 
\be
\mathcal{M}^{inv}_{\Sigma}   = \mathcal{M}^{comp,inv}_\Sigma // H^{p-1}(\p \Sigma),
\ee
is the  space of $H^{p-1}(\p \Sigma)$-invariant observables on $\mathcal{M}^{comp,inv}_\Sigma$. If we turn the $H^{p-1}(\p \Sigma)$ gauge symmetry into a global symmetry 
then the invariant phase space becomes $ \mathcal{M}^{comp,inv}_\Sigma$. Let us recall that the origin of the $H^{p-1}(\p \Sigma)$ gauge symmetry was a subgroup of the gauge transformations $G^\Sigma$ on the boundary $\p \Sigma$. Hence, we can say that the symplectic extension of the $\mathcal{M}^{inv}_{\Sigma} $ to a bigger phase space $\mathcal{M}^{comp,inv}_\Sigma$ is done by turning the boundary gauge symmetry into a global symmetry.

Let us also recall that the symplectic reduction approach is well developed only in case of the finite-dimensional spaces, while we used it for the infinite-dimensional reduction. In case of BF theory phase space  for the surface with no boundary in section  \ref{sect_invariant_BF} we observed  certain nice properties, so checking this properties for  $\mathcal{M}^{inv}_{\Sigma}$ can be considered as a consistency check. 
\begin{itemize}
\item {\it Finite dimensional phase space}. Our expression for the invariant phase space  (\ref{BF_inv})  is direct sum of the liner map kernels, what makes in a natural subspace of the direct sum of the
corresponding domains $H^{d-p}(\Sigma) \oplus H^{p}(\Sigma) $. The cohomology groups $H^{p}(\Sigma) $ are know to be finite-dimensional  so the $\mathcal{M}^{inv}_{\Sigma}  $ is finite-dimansional as well.  Moreover we can use the long exact sequences (\ref{long_exact_edge}) and (\ref{long_relative}) to prove  that the symplectic extensions $\mathcal{M}^{edge,inv}_\Sigma $ and $\mathcal{M}^{comp,inv}_\Sigma $ are finite-dimensional as well.

\item {\it Topological theory}. Our expression for the invariant phase space  (\ref{BF_inv})  is direct sum of the kernels of the linear maps between the cohomology groups. The cohomology  groups $H^{p}(\Sigma) $ are known to be topologically invariant, so the kernel of the linear map between them is also a topologically invariant.

\item {\it Self-dual} :  Our expression (\ref{BF_inv}) is manifestly invariant under the  $p \to d-p$ duality.

\end{itemize}
Let us observe that the two extensions $\mathcal{M}^{edge,inv}_\Sigma $ and $\mathcal{M}^{comp,inv}_\Sigma $ are related by duality transformation
\be\label{duality_edge}
\mathcal{M}_\Sigma^{(p)\; comp, inv}  = \mathcal{M}_\Sigma^{(d-p)\; edge, inv} 
\ee
Indeed the the $p$-form BF theory compact invariant phase space 
 \be
 \mathcal{M}_\Sigma^{(p)\; comp, inv} = H_c^p(\Sigma) \oplus H_c^p(\Sigma)^\ast  = H_c^p(\Sigma) \oplus H^{d-p}(\Sigma)
 \ee
and  the edge mode $d-p$-form BF theory compact invariant phase space  
\be
\mathcal{M}_\Sigma^{(d-p)\; edge, inv} = H^{d-p}(\Sigma) \oplus H^{d-p}(\Sigma)^\ast =H^{d-p}(\Sigma) \oplus H^{p}(i_{\p\Sigma}) 
\ee
are identical due to the cohomology relation (\ref{cohomol_same}).

\section{Conclusion}

We used the infinite-dimensional generalization of a symplectic reduction to describe the gauge-invariant phase space for BF theory. In absence of boundary  our expression for invariant phase space is identical to the one obtained by covariant phase space formalism. The gauge-invariant phase space is a direct sum of the two cohomology groups, what  is compatible with BF theory being topological theory. 

 In case surface with boundary we proposed a generalization of symplectic reduction construction for the invariant phase space. The invariant phase space preserves the topological features of BF theory as well as the $p\to d-p$ symmetry invariance. The symplectic reduction for Bf theory can be done in several steps, so that the intermediate phase spaces can be identified with the edge mode and asymptotic symmetry constructions.

Our choice of Bf theory as a prime example allowed us to provide an explicit description of various phase spaces  in terms of well known objects from algebraic topology: differential forms and de Rham cohomology. We hope that such explicit description could be   useful to check various conjectures and statements about the edge modes and asymptotic symmetries. In particular the symplectic space gluing and  TQFT description, which we briefly outline below.

\subsection{Symplectic space gluing}
Let us consider  two surfaces  $\Sigma_L$ and $\Sigma_R$ with common boundary 
\be
C = \p \Sigma_L  = - \p \Sigma_R,
\ee 
where the minus sign stands for the orientation change.   We can glue them together over the common boundary into new surface $\Sigma$
\be
\Sigma  = \Sigma_L \cup_C \Sigma_R.
\ee
There is natural question:
\begin{center} {\it 
What is the relation between  $\mathcal{M}_{\Sigma_L}$, $\mathcal{M}_{\Sigma_R}$  and $\mathcal{M}_{\Sigma} $ ? }
 \end{center}
The relation was conjectured by Donnelly and Friedel \cite{Donnelly:2016auv} to be 
\be\label{glue_conj}
\mathcal{M}_\Sigma = (\mathcal{M}_{\Sigma_L} \times \mathcal{M}_{\Sigma_R})// G_C,
\ee
with $G_C$ being some kind of diagonal  action of the  symmetry group  of  edge modes on  $\mathcal{M}_{\Sigma_L}$ and  $\mathcal{M}_{\Sigma_R}$.

There are several reasons why our analysis of BF theory can be useful in verifying this conjecture:
\begin{itemize}
\item The invariant phase spaces of BF theory are finite dimensional, so we can use finite-dimensional symplectic reduction.
\item The phase space has explicit description in terms of de Rham cohomology groups, which are well known algebraic topology objects. 
\end{itemize} 
Using all these ideas we are working on gluing conjecture verification for BF theory.  

\subsection{Extended TQFT}
The quantum version of the gluing conjecture (\ref{glue_conj}) relates the corresponding Hilbert spaces through entangled product
\be
 \mathcal{H}(\Sigma_L \cup_C \Sigma_R) = \mathcal{H}(\Sigma_L) \otimes_{G_C} \mathcal{H}(\Sigma_R).
 \ee
 In extended TQFT  there is a similar gluening axiom \cite{Carqueville:2017fmn, Freed_TQFT}  in the form of the tensor product  over algebra $A_C$ of  the right $A_{C}$-module $\mathcal{H}(\Sigma_L)$ left $A_{C}$-module $\mathcal{H}(\Sigma_R)$
\be
 \mathcal{H}(\Sigma_L \cup_C \Sigma_R) = \mathcal{H}(\Sigma_L) \otimes_{A_C} \mathcal{H}(\Sigma_R) \subset  \mathcal{H}(\Sigma_L) \otimes_{\mathbb{C}} \mathcal{H}(\Sigma_R).
\ee
The two gluing operations become the identical if we  conjecture that the algebra $A_C$ is the group algebra  of $G_C$
\be\label{etqft_match}
 A_C = \mathbb{C}[G_C] = \left\{ \sum_{i=1}^{|G|} \lambda_i g_i \;|\; \lambda_i \in \mathbb{C}\right\}.
\ee
The consistency of ETQFT among other conditions require that 
\be\label{tqft_cons}
A_{C} =  \mathcal{H}(C\times[0,1]).
\ee
The form of $A_C$ is fixed from our conjecture (\ref{etqft_match}) so given our explicit form  $\mathcal{M}_\Sigma^{inv}$ we can check the consistency  (\ref{tqft_cons}) and  analyze remaining 
extended TQFT axioms.

\section*{Acknowledgments}
V.L. is grateful to  Yasha Neiman and Sudip Ghosh  for useful conversations. This work  is supported by the Quantum Gravity Unit of the Okinawa Institute of Science and Technology Graduate University (OIST), Japan.
\appendix

\bibliography{symp_red_ref}{}
\bibliographystyle{utphys}

\end{document}